\documentclass[12pt]{iopart}
\usepackage{amssymb}
\nopagebreak
\begin{document}

\nopagebreak

\title{Critical exponents for Gaussian fixed point of renormalization} 

\author{Witold Haliniak$^1$ and Wojciech Wislicki$^2$}

\address{$^1$ University of Warsaw, Faculty of Mathematics, Informatics and Mechanics, Banacha~2, PL-02-097 Warszawa\\

$^2$ A. Soltan Institute for Nuclear Studies, Laboratory for High Energy Physics, Hoza~69, PL-00-681 Warszawa
}

\ead{$^!$ \mailto{witeha@gmail.com}, $^2$ wislicki@fuw.edu.pl}

\begin{abstract}
We present mathematical details of derivation of the critical exponents for the free energy and magnetization in the vicinity of the Gaussian fixed point of renormalization.
We treat the problem in general terms and do not refer to particular models of interaction energy.
We discuss the case of arbitrary dispersion of the fixed point.

\vspace{5mm}



\end{abstract}

\maketitle

\section{Introduction}

The method of renormalization group (RG), representing nowadays a mature tool for treatment of critical phenomena and multi-scale systems, attracted also attention due to its interesting mathematical connections.
One of them was the recognition that strictly $\alpha$-stable random variables \cite{gnedenko} are fixed points of the RG \cite{jonalasinio_1}.
These results appeared to be very useful when considering scaling properties of thermodynamic potentials near the critical points of systems with many degrees of freedom.
Exploiting the concept of the generalized homogeneous function (GHF) and its scaling properties one can derive general expressions for critical exponents \cite{hankeystanley}.
They can be explicitly calculated in the vicinity of the Gaussian fixed points.
The results, or hints leading to them, are spread over numerous papers and treated at different levels of certainty (cf. ref. \cite{jonalasinio_2} and discussions therein).
In this note we aim to discuss some details and subtleties of such derivations with mathematical rigour which, in our opinion, is often lacking or insufficiently accounted in the literature.
After introducing, in chapter 2, basic notions with necessary clarifications, we present our discussions in chapter 3.

\section{Basic notions}

\subsection{Renormalization and $\alpha$-stable random variables}

Assume that $\mathbf X=\{X_i\}_{i=-\infty}^\infty$ represents a sequence of real, independent and identically distributed ({\it iid}) random variables and $\rho(x)$ is their density.
Let $\{T_n\}_{n\ge 1}$ be a one-parameter family of the renormalization operators on $\mathbf X$ defined as
\begin{eqnarray}
\big(T_n(\mathbf X)\big)_i=\frac{1}{n^\delta}\sum_{j=in}^{(i+1)n-1}X_j, \quad\quad\quad \delta>0.
\label{eq1}
\end{eqnarray}
This family with an operation $T_n\circ T_m(.)=T_n(T_m(.))=T_{nm}(.)$ constitutes the RG.
For brevity, we use notation $T_n\, \rho(x)$ for density of renormalized random variable.

A sequence of the {\it iid} stricly $\alpha$-stable random variables is a fixed point of renormalization with $\delta=1/\alpha$ ($0<\alpha\le 2$) \cite{samorodnitsky}.
Densities of such variables appear as the limit of sequences of densities resulting from successive applications of $T_n$ with fixed $n$.
In this sense, for given $\alpha$, all variables with asymptotic behaviour of densities the same as strictly $\alpha$-stable 
\begin{eqnarray}
\lim_{x\rightarrow\pm\infty}|x|^{1+\alpha}\rho(x)=\mbox{const}\ne 0, \quad\quad\quad \alpha<2
\label{eq2}
\end{eqnarray}
are called to belong to the {\it domain of attraction} of this stricly $\alpha$-stable variable.
In case of $\alpha=2$, random variables with finite variance are attracted by the normal variable which corresponds to the central limit theorem (CLT).

Consider density $\rho(x)$ in the vicinity of the fixed point and denote by $\rho^\ast(x)$ the density of strictly $\alpha$-stable random variable being the fixed point of renormalization.
We consider the case of $n=2$.
We denote small deviation from the fixed point (in any functional norm, e.g. $L^2$) by $\eta(x)=\rho(x)-\rho^\ast(x)$ and renormalize $\eta$
\begin{eqnarray}
\eta^\prime(x) & = & DT_2\eta(x) \nonumber \\ 
               & = & T_2(\rho^\ast+\eta)(x)-T_2\rho^\ast(x).
\label{eq4}
\end{eqnarray}
For small $\eta$ we linearize $DT_2 \eta$ by neglecting ${\mathcal O}(\eta^2)$ terms and arrive to its explicit form
\begin{eqnarray}
DT_2\eta(x)=2^{1/\alpha+1}\int_{-\infty}^\infty dy\,\rho^\ast(2^{1/\alpha}x-y)\eta(y).
\end{eqnarray}

If $\phi_n(x)$ stand for eigenfunctions of $DT_2$ and $\lambda_n$ for its eigenvalues then 
the $\eta(x)$ and $\eta^\prime(x)$ can be expanded in basis $\{\phi_n(x)\}_{n=1}^\infty$:
\begin{eqnarray}
\eta^{(\prime)}(x)=\sum_{n=1}^\infty v_n^{(\prime)}\phi_n(x)
\label{eq6}
\end{eqnarray}
and the coefficients $\{v_n\}_{n=0}^\infty$ transform covariantly
\begin{eqnarray}
v_n^\prime=\lambda_n\,v_n.
\label{eq7}
\end{eqnarray}
Assume that the random variable $\mathbf X$ belongs to the normal domain of attraction of variable $Y$.
This means \cite{gnedenko} that
\begin{eqnarray}
\forall_{n\ge 1}\;\;\; \exists_{a_n,b_n\in\mathbb{R}}\;\;\; \frac{1}{a_n}(X_1 +\ldots +X_n)+b_n \stackrel{d}{\longrightarrow} Y,
\label{eq7p}
\end{eqnarray}
where $a_n=n^{1/\alpha}$.
If $Y$ (with density $\rho^\ast$) is the fixed point of renormalization $T_2$, then
\begin{eqnarray}
\underbrace{T_2\circ\ldots\circ T_2}_{\mbox{\scriptsize n times}}(\mathbf X) & = & T_{2^n}(\mathbf X)\stackrel{d}{\longrightarrow} Y.
\label{eq8}
\end{eqnarray}
The $\stackrel{d}{\longrightarrow}$ stands for the convergence in the sense of distribution.
In such case $\lambda_n<1$, unless $v_n\ne 0$, and those $v_n$ are called {\it irrelevant parameters}.
The parameters $v_n$ for which $\lambda_n>1$ are called {\it relevant} and those corresponding to $\lambda_n=1$ are called {\it marginal}.
The subspace of $\{v_n\}_{n=1}^\infty$ consisting of relevant parameters is called a {\it critical surface} and is identical to the domain of attraction of stricly stable variable $Y$ with density $\rho^\ast$.

We consider random variable close to the domain of attraction of strictly $\alpha$-stable variable but not belonging to it.
We further consider $\alpha=2$ with a finite variance $\sigma^2$.
In this case the eigenfuntions are given as (cf. e.g. \cite{jonalasinio_2} or \cite{honerkamp})
\begin{eqnarray}
\phi_n^\sigma(x)=\frac{1}{\sqrt{2\pi}\sigma}e^{-x^2/2\sigma^2}H_n(x/\sigma),\quad\quad n=0,1,2,\ldots 
\label{eq9}
\end{eqnarray}
where $H_n$ stands for the Hermite polynomials and the eigenvalues are equal to
\begin{eqnarray}
\lambda_n^\sigma=2^{1-n/2}\sigma^n.
\label{eq11}
\end{eqnarray}
As seen from eq.~(\ref{eq11}), the number of relevant parameters is determined by the condition $\sigma>2^{1/2-1/n}$.
We assume there exists at least one relevant parametr which is equivalent to the condition $\sigma>1/\sqrt{2}$.

\subsection{Scaling and critical exponents}

The notion of {\it critical exponents} can be introduced in an abstract way for any function $h:{\mathbb R}\rightarrow {\mathbb R}^+$ continuous around $0$.
If there exist limits
\begin{eqnarray}
\gamma^\pm = \lim_{\varepsilon\rightarrow 0^\pm}\frac{\ln h(\varepsilon)}{\ln (\pm\varepsilon)}
\label{eq11a}
\end{eqnarray}
then we call $\gamma^\pm$ the critical exponents of $h$.
The function $h$ then exhibits a power-law behaviour around 0:
\begin{eqnarray}
h(\varepsilon) \sim \pm\varepsilon^{\gamma^\pm}, \quad\quad \varepsilon\rightarrow 0^\pm.
\label{eq11b}
\end{eqnarray}

A function $g(x_1,x_2)$ is called the GHF \cite{hankeystanley} if there exist three numbers $a_1, a_2, a_g$, at least one of them non-zero, such that for all positive $\lambda$ 
\begin{eqnarray}
g(\lambda^{a_1}x_1,\lambda^{a_2}x_2)=\lambda^{a_g}g(x_1,x_2).
\label{eq12}
\end{eqnarray}
Only two of the parameters $a_1, a_2, a_g$ are independent, as all of them can be rescaled by a common factor.
If $a_g\ne 0$ then the parameters can be rescaled by $1/a_g$ and eq. (\ref{eq12}) reads
\begin{eqnarray}
g(\bar\lambda^{a_1/a_g}x_1,\bar\lambda^{a_2/a_g}x_2)=\bar\lambda\, g(x_1,x_2),
\label{eq13}
\end{eqnarray}
where $\bar\lambda=\lambda^{a_g}$.
In case of $a_g=0$ the GHF is called {\it scale invariant}.
If $g$ is GHF then its all partial derivatives are GHFs.

The GHF's scaling parametrs can be related to their critical exponents.
Assuming $|x_{1(2)}|\lambda^{a_{1(2)}}=1$ and $x_{2(1)}=0$ one arrives to 
\begin{eqnarray}
g(x_1,0) & = & |x_1|^{a_g/a_1}g\big({\rm sgn} (x_1),0\big)
\label{eq15}
\end{eqnarray}
and analogously for $g(0,x_2)$.
The quantities $\alpha_{g_2}=a_g/a_1$ and $\alpha_{g_1}=a_g/a_2$ are the critical exponents of the GHF g.
Using eq.~(\ref{eq15}), critical exponents $\alpha_{g_{1(2)}}$ can be derived without even knowing the values of scaling parameters $a_1, a_2, a_g$.
It is sufficient to know $g(1,0)$ and $g(\bar x_1,0)$ for some $\bar x_1>0$ (or $g(0,1)$ and $g(0,\bar x_2)$ for some $\bar x_2>0$, respectively) and then
\begin{eqnarray}
\alpha_{g_1} & = & \frac{\ln g(0,\bar x_2)/g(0,1)}{\ln|\bar x_2|}
\label{eq15a}
\end{eqnarray}
and $\alpha_{g_2}$ by analogy.
The quantities $g(\bar x_1,0)/g(1,0)$ and $g(0,\bar x_2)/g(0,1)$ are positive because the numerators and denominators have the same signs provided eq.~(\ref{eq15}) is fulfilled.
Critical exponents for the derivative of GHF are discussed in detail in chapt.~3.

Relation of such introduced critical exponents to those known from thermodynamics is given by scaling properties of thermodynamic potentials.
For a random variable with density $\rho$ we introduce the {\it free energy} $F$:
\begin{eqnarray}
F[\rho,t] =\ln\int_{-\infty}^{\infty}dx\,e^{tx}\rho(x)
\label{eq16}
\end{eqnarray}
which transforms under renormalization $T_2$ of strictly $\alpha$-stable random variable
\begin{eqnarray}
F[\rho^\prime(x),t]=2F[\rho(x),2^{-1/\alpha}t].
\label{eq17}
\end{eqnarray}
Using expansion (\ref{eq6}) and transformation of coefficients (\ref{eq7}) this can be rewritten for infinite number of variables and in terms of $v_n$s as
\begin{eqnarray}
F(\lambda_1v_1, \lambda_2v_2,\ldots,2^{1/\alpha}t) = 2F(v_1, v_2,\ldots,t).
\label{eq18}
\end{eqnarray}
In the following we assume that close to the critical point the free energy decomposes into the sum of the regular and the singular parts.
For the singular part, for which we hereon use the symbol $F$, we further assume the {\it scaling hypothesis} stating that $F$ is asymptotically GHF \cite{hankeystanley}.

\section{Derivation of the critical exponents for the Gaussian fixed point}

\subsection{Existence of relevant parameters}

Consider random variable $X$ close to the domain of attraction of the Gaussian fixed point, with expected value 0 and the variance $\sigma^2$, but not belonging to it.
From CLT it follows that any random variable with finite variance belongs to the domain of attraction of the Gaussian, so that we have to assume non-existence of the second moment of $X$.
We also assume that its density has expansion (\ref{eq6}) in terms of eigenfunctions (\ref{eq9}).
Identification of relevant parameters is only possible when the variance $\sigma^2$ is known.
If $\sigma>\sqrt{2}$ then the sequence $\{\lambda_n^\sigma\}_{n=1}^\infty$ (\ref{eq11}) is increasing and there is infinitely many relevant paramenters.
In case of $\sigma<\sqrt{2}$ the sequence $\{\lambda_n^\sigma\}_{n=1}^\infty$ is decreasing to zero and the largest relevant eigenvalue is $\lambda_1^\sigma=\sqrt{2}\sigma$.
If $1/\sqrt{2}<\sigma<\sqrt{2}$ then the number of relevant parameters is finite.
If there exists at least one relevant parameter then this is $v_1$ and therefore it is sufficient to consider only critical exponents in dimension $v_1$.

\subsection{Free energy and its first derivatives in the vicinity of Gaussian fixed point}

Extend from the scaling condition for $F$ written for any $\lambda>0$ in the form
\begin{eqnarray}
F(\lambda^{a_1}v_1,\lambda^{a_2}v_2,\ldots,\lambda^{a_t}t)=\lambda^{a_f}F(v_1,v_2,\ldots,t)
\label{eq19}
\end{eqnarray}
and take its derivative over $v_k$, and denote $m_k=\partial F/\partial v_k$, $k\ge 1$.
Direct determination of critical exponents for $F$ and $m_k$ requires putting all but one arguments equal to zero.
This method cannot be applied in our case because both $F$ and $m_k$ vanish identically for $t=0$.
In order to see this we write down the free energy using explicit expansion:
\begin{eqnarray}
F(v_1,v_2,\ldots,t)=\ln (I_0+\sum_{n=1}^\infty v_n I_n),
\label{eq20}
\end{eqnarray}
where
\begin{eqnarray}
I_n=\left\{ \begin{array}{ll}
                       e^{\sigma^2t^2/2}, &  \quad\quad n=0 \\
                       \int_{-\infty}^\infty dx\,e^{tx}\phi_n^\sigma (x), & \quad\quad n>0
                       \end{array}
                     \right.
\label{eq21}
\end{eqnarray}
Assuming homogeneous convergence of the series $\sum_{n=1}^\infty v_n\phi_n^\sigma e^{tx}$ on $\mathbb R$ for all $t$ one interchanges integration and summation in eq.~(\ref{eq20}).
Homogeneous convergence of the series $\sum_{n=1}^\infty v_nH_n(x)$ follows further from the Cauchy condition and boundedness of $e^{t_0x-x^2/2\sigma^2}$ for any $t_0$.
Homogeneous convergence implies convergence of each term to 0, for any $t$, and then:
\begin{eqnarray}
\lim_{n\rightarrow\infty}v_nH_n(x/\sigma)=0,\quad\quad\quad \mbox{for any}\; x.
\label{eq22}
\end{eqnarray}
The highest order term for the $n$-th Hermite polynomial $H_n(x/\sigma)$ is equal to $(x/\sigma)^n$.
For sufficiently large $x$ there can be $x/\sigma>1$ and therefore the series $\{v_n\}_{n=1}^\infty$ must converge at least exponentially.
All $I_n$ can thus be explicitly calculated such that
\begin{eqnarray}
F(v_1,v_2,\ldots,t)=\ln e^{\sigma^2t^2/2}(1+\sum_{n=1}^\infty v_n (\sigma t)^n)
\label{eq23}
\end{eqnarray}
and
\begin{eqnarray}
m_k(v_1,v_2,\ldots,t)=\frac{(\sigma t)^k}{1+\sum_{n=1}^\infty v_n (\sigma t)^n},\quad\quad k\ge 1.
\label{eq24}
\end{eqnarray}
It follows from eqs. (\ref{eq23}, \ref{eq24}) that
\begin{eqnarray}
F(v_1,v_2,\ldots,0)=m_k(v_1,v_2,\ldots,0)=0, \quad\quad k\ge 1.
\label{eq25}
\end{eqnarray}

\subsection{Derivation of the final formulae}

We put $v_k=0$ for $k>1$ and use an abbreviation $M_1(v_1,t)=m_1(v_1,0,\ldots,t)$.
Then
\begin{eqnarray}
M_1(\lambda^{a_1}v_1,\lambda^{a_t}t)=\lambda^{a_f-a_1}M_1(v_1,t).
\label{eq26}
\end{eqnarray}
Having the scaling condition valid for any $\lambda>0$ we assume $\lambda=|v_1|^{-1/a_1}$ and obtain
\begin{eqnarray}
M_1(v_1,t)=|v_1|^{a_f/a_1-1}M_1({\rm sgn}\, v_1,|v_1|^{-a_t/a_1}).
\label{eq27}
\end{eqnarray}
In order to calculate the critical exponent $\alpha_{M_1}$ we take $M_1$ at some $t_0\ne 0$ and use definition (\ref{eq11a}). Hence
\begin{eqnarray}
\alpha_{M_1} & = & \frac{a_f}{a_1}-1+L.
\label{eq28}
\end{eqnarray}
The term $L$ has to be treated with caution.
If $a_t/a_1<0$ then using eq.~(\ref{eq24}) one arrives to 
\begin{eqnarray}
L & = & -\frac{a_t}{a_1}+\lim_{v_1\rightarrow 0^+}\frac{\sigma t_0 \frac{a_t}{a_1}}{|v_1|^{a_t/a_1}+\sigma t_0} \nonumber \\
  & = & -\frac{a_t}{a_1}.
\label{eq29}
\end{eqnarray}
For $a_t/a_1>0$ we have $L=0$.
Summarizing:
\begin{eqnarray}
\alpha_{M_1}=\left\{ \begin{array}{ll}
                       \frac{a_f}{a_1}-1, &  \quad\quad \frac{a_t}{a_1}>0 \\
                       \frac{a_f-a_t}{a_1}-1, & \quad\quad \frac{a_t}{a_1}<0.
                       \end{array}
                     \right.
\label{eq30}
\end{eqnarray}
The $\alpha_{M_1}$ does not depend on $t_0$, as expected.

Using the scaling condition (\ref{eq18}) with eigenvalues (\ref{eq11}), combinig it with the scaling condition in form of eq.~(\ref{eq12}) and substituting $\lambda=\bar\lambda^p$ ($\bar\lambda>0$ and $p\in\mathbb R$) we get
\begin{eqnarray}
pa_k & = & 1+k\frac{\ln \sigma/\sqrt{2}}{\ln 2},\quad\quad k\ge 1, \nonumber \\
pa_t & = & \frac{1}{2}, \nonumber \\
pa_f & = & 1.
\label{eq31}
\end{eqnarray}
From this follows that $a_t\ne 0$ and therefore we drop the case $\frac{a_t}{a_1}=0$ in eq.~(\ref{eq30}).

We can finally write down the formulae for the critical exponents of $F$ and $M_1$, denoted by $\alpha_{F_1}$ and $\alpha_{M_1}$
\begin{eqnarray}
\alpha_{F_1} & = & \frac{1}{1+\ln(\sigma/\sqrt{2})/\ln 2},
\label{eq32}
\end{eqnarray}
\begin{eqnarray}
\alpha_{M_1}=\left\{ \begin{array}{ll}
                       \frac{1}{1+\ln(\sigma/\sqrt{2})/\ln 2}-1, &  \quad\quad \frac{a_t}{a_1}>0 \\
                       \frac{1/2}{1+\ln(\sigma/\sqrt{2})/\ln 2}-1, & \quad\quad \frac{a_t}{a_1}<0.
                       \end{array}
                     \right.
\label{eq33}
\end{eqnarray}
Since $a_t/a_1>0$ for $\sigma>1/\sqrt{2}$, then the second case in eq.~(\ref{eq33}) can be ignored provided the number of relevant parameters is finite.

\end{document}